\documentclass[
    reprint,
    amsmath,
    amssymb,
    aps,
    prl,
    nobibnotes,
    superscriptaddress
]{revtex4-2}

\usepackage{graphicx}
\usepackage{dcolumn}
\usepackage{bm}
\usepackage[colorlinks=true, citecolor=blue, linkcolor=black, urlcolor=black]{hyperref}
\usepackage{amsmath}
\usepackage{amssymb}
\usepackage{amsfonts}

\begin{document}
	
	\title{Accurate qLDPC decoding under real-time constraints}

\author{Chenzewei Qing}
\thanks{These authors contributed equally to this work.}
\affiliation{Beijing Academy of Quantum Information Sciences, Beijing 100193, China}
\affiliation{State Key Laboratory of Low Dimensional Quantum Physics, Tsinghua University, Beijing 100084, China}

\author{Tengyu Zhang}
\thanks{These authors contributed equally to this work.}
\affiliation{Beijing Academy of Quantum Information Sciences, Beijing 100193, China}

\author{Xiaoqin Luo}
\affiliation{Beijing Academy of Quantum Information Sciences, Beijing 100193, China}

\author{Xiaolu Su}
\email{suxiaolu0810@gmail.com}
\affiliation{Beijing Academy of Quantum Information Sciences, Beijing 100193, China}

\author{Li You}
\affiliation{Beijing Academy of Quantum Information Sciences, Beijing 100193, China}
\affiliation{State Key Laboratory of Low Dimensional Quantum Physics, Tsinghua University, Beijing 100084, China}
\affiliation{Frontier Science Center for Quantum Information, Beijing, China}
\affiliation{Hefei National Laboratory, Hefei, Anhui 230088, China}

	\date{\today}
	

	\begin{abstract}
		Fault-tolerant quantum computation requires reconciling the computational demands of accurate decoding with the stringent timing constraints of real-time error correction. We present Spatiotemporal Covariance-Optimized Parallel Ensemble (SCOPE),
		a physics-informed framework for real-time decoding of quantum low-density
		parity-check codes (qLDPC). By exploiting the spatial symmetries of bivariate-bicycle 
		(BB) codes and the temporal regularity of repeated syndrome extraction, SCOPE
		uncovers a spatiotemporally covariant decoding structure and organizes the
		circuit-level problem into a layered decoding strategy. We further compress
		the high-dimensional evolution of belief propagation (BP) into a low-dimensional
		dynamical representation, revealing distinct signatures of successful and
		failed decoding and enabling early detection of unpromising trajectories.
		Guided by these dynamical signatures, SCOPE explores multiple perturbed decoding trajectories in parallel, enabling early termination and reseeding within a fixed serial iteration budget, with message-passing parameters optimized offline by simulated annealing.
		Circuit-level simulations of the $[[144,12,12]]$
		BB code yield a logical error rate of
		$p_L=4.82\times10^{-10}$ at a physical error rate of $p=10^{-3}$.
		We further realize a pipelined two-arm memory-augmented BP (Mem-BP) core at the
		register-transfer level. Functional simulations at a
		\(250\,\mathrm{MHz}\) clock yield a \(16\,\mathrm{ns}\) BP
		iteration interval per arm and a mean input-to-result latency
		of \(173.8\,\mathrm{ns}\) at \(p=5\times10^{-3}\).
	\end{abstract}

	\maketitle

	Quantum computing promises to fundamentally revolutionize information processing across a broad spectrum of disciplines~\cite{troyrer17, lloyd96,farhi2014, grover1996}. Owing to the intrinsic fragility of quantum states, the performance of quantum computation is highly susceptible to noise and errors. Although the impact of certain errors can be mitigated through various techniques\cite{Wu_2022,Veps_l_inen_2020, weber2024robust}, quantum error correction (QEC) is indispensable for scalable quantum computation. Fault-tolerant quantum computation therefore requires not only reliable operations on encoded states\cite{Gottesman_1998}, but also efficient classical syndrome processing to keep pace with repeated QEC cycles and support timely feedback when needed\cite{Battistel_2023}. Recent QEC experiments across different physical platforms have demonstrated rapid progress in both error suppression and system-level implementation\cite{atom2026, Google2024, ryananderson2021}.
	
	Most demonstrations of real-time QEC to date have been based on surface codes, with differing emphases on decoding throughput
	and feedback latency. One approach prioritizes throughput, using software decoders to process syndrome streams without backlog\cite{Skoric_2023, wu2023}, as in Google Willow~\cite{Google2024}; another prioritizes feedback latency, embedding decoders in dedicated hardware for low-latency
	response~\cite{caune2024,yang2026,Barber_2025}. Although surface codes are well suited to lattice-based architectures\cite{fowler2012surface}, quantum low-density parity-check (qLDPC) codes are increasingly favored for their higher encoding rates and lower qubit overhead\cite{Tillich_2014, xu2024constant}, with recent experiments showing encouraging progress\cite{wang2026demonstration,tham2026}. Realizing these advantages in practical fault-tolerant quantum computation requires classical decoders that achieve low logical error rates while providing sufficiently low latency for time-critical feedback\cite{Skoric_2023, Battistel_2023, Hillmann_2025}.
	
	For qLDPC codes, belief propagation (BP) has become the dominant iterative decoding framework. Most existing qLDPC decoders are developed by advancing or extending BP-based methods through post-processing, modified message passing, or ensemble-based strategies\cite{Roffe_2020,muller2025,Maan_2026,Hillmann_2025, Wang_2026, wang2026gpubasedparallelbpdecoding}. In particular, Relay-BP has demonstrated an amortized average decoding time of 40 ns per QEC cycle at a physical error rate of $p=10^{-3}$\cite{maurer2025}, indicating its potential for high-throughput real-time qLDPC decoding. These results show that BP provides a powerful and flexible foundation for practical qLDPC decoder design.
	
	\begin{figure*}[t]
		\centering
		\includegraphics[width=\textwidth]{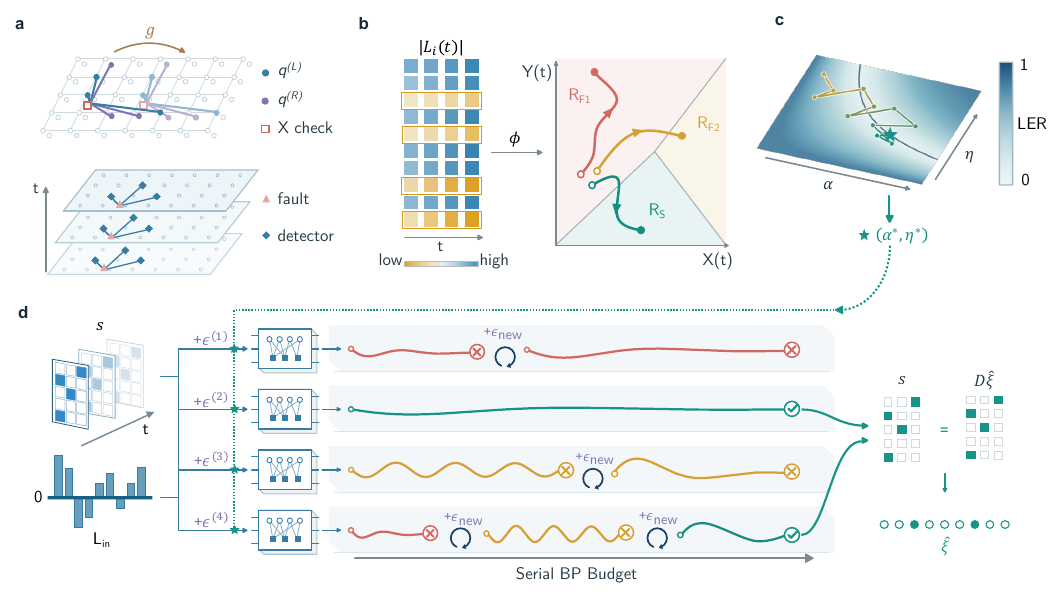}
		\caption{
			\textbf{Structure- and dynamics-guided quantum error correction with SCOPE.} \textbf{a,} Spatiotemporal covariance of the circuit-level decoding operator. Top, spatial covariance: The quasi-cyclic structure of BB codes preserves the connectivity between stabilizer checks and data qubits under a common cyclic translation $g$ of their spatial coordinates. Bottom, temporal covariance: Repeated syndrome extraction ensures that faults of the same type at corresponding circuit locations in different rounds generate time-shifted detector-response patterns with unchanged relative structure, away from temporal boundaries. \textbf{b}, High-dimensional BP dynamics are characterized by low-confidence log-likelihood-ratio (LLR) statistics extracted from recent iterations and projected into a two-dimensional diagnostic space. The resulting representation distinguishes the successful decoding region ($R_S$) from two characteristic failure regions ($R_{F1}$ and $R_{F2}$), enabling early identification of unpromising trajectories. \textbf{c}, Simulated annealing is used to optimize the Min-Sum message-weighting parameters $(\alpha,\eta)$ offline. Monte Carlo estimates of the logical error rate (LER) guide the search toward the optimized parameters $(\alpha^*,\eta^*)$. In the schematic parameter map, equal-color bands follow hyperbolic contours of constant \(\alpha\cdot\eta\). \textbf{d}, Overall workflow of SCOPE's parallel decoding. Four concurrent BP arms process a common detector syndrome $\mathbf{s}$ and input LLR profile $\mathbf{L}_{\mathrm{in}}$, sharing the optimized parameters from \textbf{c} while applying independent random perturbations to their input LLRs. The dynamical criteria introduced in \textbf{b} identify unpromising trajectories for early termination, allowing each arm to restart from the common input with a fresh perturbation using its remaining iteration budget. The preceding BP preprocessing and final OSD fallback are omitted for clarity.}
		\label{fig:fig1}
	\end{figure*}

	Despite these advances, achieving low logical error rates within a bounded decoding-latency budget remains challenging for three closely related reasons. First, at the physical-system level, fault-tolerant QEC requires repeated syndrome extraction in the presence of measurement and circuit-level faults, producing correlated spacetime syndrome data that must be processed within a stringent QEC cycle-time budget\cite{Battistel_2023, Bravyi2024}. Second, iterative qLDPC decoding exhibits complex dynamical behavior. Although several failure mechanisms, including degeneracy-related locking, trapping-set effects, and symmetry-induced non-convergence, have been identified\cite{poulin2008, Raveendran_2021, fregona2026}, how the evolution of BP beliefs distinguishes successful convergence from persistent failure remains less well understood. Third, practical real-time deployment is constrained by two implementation bottlenecks: BP performance depends sensitively on message-passing parameter choices\cite{muller2025}, while hard decoding instances can produce long serial iteration
	tails\cite{Wang_2026}, making it difficult to meet a fixed
	feedback-time budget\cite{Battistel_2023}. Addressing these intertwined physical, dynamical, and implementation challenges is crucial for scalable, high-performance qLDPC decoding under real-time constraints.
	
	Here, we propose SCOPE as a highly parallel and accurate decoder that addresses these three challenges (Fig. \ref{fig:fig1}). First, we exploit the temporal and spatial covariance of the spacetime decoding matrix to develop a layered decoding strategy, substantially reducing decoding latency and computational overhead. Second, we develop a statistical description of low-confidence decoding defects, which provides a low-dimensional representation of iterative decoding dynamics and enables early identification of unpromising trajectories, thus improving iteration efficiency. Third, we combine simulated-annealing-based optimization of Min-Sum BP parameters with a parallel multi-arm architecture, automating parameter calibration and reducing serial decoding depth under real-time constraints. Numerical simulations on the $[[144,12,12]]$ bivariate bicycle (BB) code demonstrate state-of-the-art performance in both decoding accuracy and latency.
	
	\subsection*{Spatiotemporal covariance of the decoding operator}
	
	We consider the circuit-level decoding problem
	$D\boldsymbol{\xi}=\mathbf{s}$, where
	$D\in\mathbb{F}_2^{M\times N}$ is the decoding matrix,
	$\boldsymbol{\xi}\in\mathbb{F}_2^{N}$ specifies the effective
	circuit-fault configuration, and
	$\mathbf{s}\in\mathbb{F}_2^{M}$ is the measured detector syndrome.
	Each column of $D$ therefore represents the detector signature shared by a set of equivalent elementary circuit faults, while each row corresponds to a detector
	outcome.
	
	For a BB code, spatial locations are indexed
    by the direct product of cyclic groups
	$\mathcal{M}=\mathbb{Z}_{\ell}\times\mathbb{Z}_{m}$. We denote the
	spatial coordinate of a fault by $\alpha\in\mathcal{M}$ and that of a
	detector by $\beta\in\mathcal{M}$. For fixed fault and detector types,
	whose labels are suppressed below, a matrix element is therefore written
	as $D_{(t',\beta),(t,\alpha)}$. The quasi-cyclic structure of the BB
	code implies that simultaneously translating the fault and detector by
	the same $g\in\mathcal{M}$ leaves their relation unchanged:
	\begin{equation}
		D_{(t',\beta+g),(t,\alpha+g)}
		=
		D_{(t',\beta),(t,\alpha)} .
		\label{eq:spatial_covariance}
	\end{equation}
	Thus, the spatial dependence of the decoding operator is determined by
	the relative displacement $\beta-\alpha$, rather than by the absolute
	positions of the fault and detector.
	
	An analogous covariance arises in time from the repetition of the same
	syndrome-extraction circuit. Within the repeated bulk of the measurement
	sequence, simultaneously shifting the fault and detector rounds by
	$\Delta$ preserves their circuit-level relation:
	\begin{equation}
		D_{(t'+\Delta,\beta),(t+\Delta,\alpha)}
		=
		D_{(t',\beta),(t,\alpha)} ,
		\label{eq:temporal_covariance}
	\end{equation}
	provided that both configurations remain away from the temporal
	boundaries. Hence, in the bulk, the temporal dependence is determined
	only by the relative round separation $t'-t$.
	
	Combining the two translation covariances illustrated in Fig. \ref{fig:fig1}a, the spacetime decoding matrix
	can be written in the form
	\begin{equation}
		D_{t',t}\simeq K_{t'-t},
		\qquad
		(K_{\Delta t})_{\beta+g,\alpha+g}
		=
		(K_{\Delta t})_{\beta,\alpha},
		\label{eq:spatiotemporal_covariance}
	\end{equation}
	where $\Delta t=t'-t$ and $K_{\Delta t}$ denotes the fault--detector
	coupling block at temporal separation $\Delta t$. Here, $\simeq$
	reflects finite-window boundary effects: the relation is exact in the
	repeated bulk but is modified in the initial and final rounds, including
	the final data-syndrome contribution.
	
	Causality further requires $K_{\Delta t}=0$ for $\Delta t<0$, since a
	fault cannot affect a detector in an earlier round. Because the detector response to a localized circuit fault has finite
	temporal support, there is also a finite
	range $w$ beyond which $K_{\Delta t}=0$. The circuit-level decoding
	operator is therefore block-Toeplitz-like in time, quasi-cyclic within
	each spatial block, and causal with finite temporal support.
	
\begin{figure}
		\centering
		\includegraphics[width=\columnwidth]{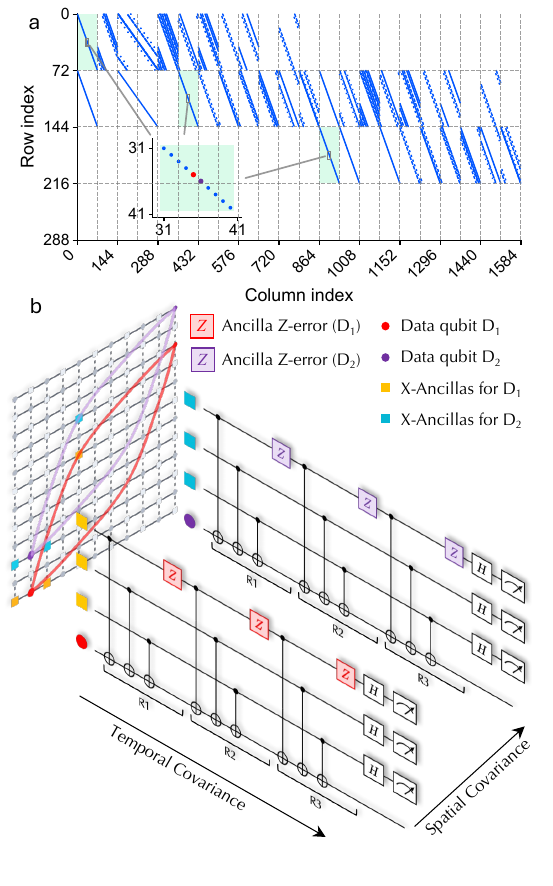}
		\caption{\textbf{Spatiotemporal structure of the circuit-level decoding
				operator.}
			\textbf{a}, Representative sector of the circuit-level decoding matrix for
			the $[[144,12,12]]$ BB code. Blue pixels denote nonzero entries; the inset
			shows a representative block with selected entries highlighted in red and
			purple.
			\textbf{b}, Corresponding circuit-level fault locations and detector
			responses. $D_1$ and $D_2$ denote adjacent data-qubit locations, while
			$R_1$--$R_3$ label successive syndrome-extraction rounds.}
		\label{matrix}
	\end{figure}

	The resulting structure is illustrated in Fig.~\ref{matrix}a for the
	$[[144,12,12]]$ BB code. The circuit-level decoding matrix exhibits
	repeated coupling patterns along the temporal block diagonals, while
	translated fault locations generate corresponding shifted patterns within
	each spatial block. These two structures arise directly from repetition
	of the same syndrome-extraction circuit across measurement rounds and
	from the quasi-cyclic spatial connectivity of the BB code, respectively.
	Representative fault locations and their detector responses are shown in
	Fig.~\ref{matrix}b. The circuit-level construction of $D$ is described in Methods, while
	derivations of the spatial and temporal covariance relations, including
	finite-window boundary effects, are provided in Supplementary Information.
	
	\subsection*{Low-dimensional representation of BP dynamics}
	
	Internal reliability and oscillation information has been used to guide recovery from decoding failures\cite{Wang_2026, taghipour2026, hu2026fastaccurateneuraltriage}. Convergence information has also been used to estimate decoding
	confidence for post-selection\cite{aydin2026softdecodingquantumldpc}. Here, we track the evolution of the decoder belief state to construct a low-dimensional diagnostic representation. At iteration $t$, the state is represented by the posterior log-likelihood-ratio (LLR) vector
	\begin{equation}
		\mathbf{L}(t)
		=
		\big(
		L_1(t),L_2(t),\ldots,L_N(t)
		\big)
		\in \mathbb{R}^{N},
		\nonumber
	\end{equation}
	where $L_i(t)$ denotes the posterior LLR of variable node $i$. Belief propagation therefore generates a trajectory in an
	$N$-dimensional belief space. We find that the distinction between
	convergent and non-convergent decoding attempts is strongly reflected
	in the evolution of the low-confidence part of the LLR distribution,
	suggesting that the relevant high-dimensional dynamics can be described
	through a small number of collective observables.
	
	Writing $\ell_i(t)=|L_i(t)|$ for the reliability of variable $i$, we identify low-confidence variables through $I_i(t)=\mathbf{1}\!\left[\ell_i(t)<\tau\right]$, where $\tau$ is a fixed reliability threshold. We then define the zeroth- and first-order low-confidence moments as
	\begin{equation}
		M_0(t)
		=
		\sum_i I_i(t),
		\quad
		M_1(t)
		=
		\sum_i
		I_i(t)
		\big[
		\tau-\ell_i(t)
		\big].
	\end{equation}
	Here, $M_0$ measures the population of low-confidence variables, whereas $M_1$ quantifies their aggregate severity. These two quantities provide an interpretable coarse graining of the microscopic LLR state and form the principal observables used to characterize its late-stage dynamics.
	
	To distinguish different dynamical behaviors, we define a projection: $\Phi:(\mathbb{R}^{N})^{h}\longrightarrow
	\mathbb{R}^{2}$, where $h$ denotes the temporal window used to characterize the decoding trajectory. Specifically,
	\begin{equation}
		\Phi\!\left(
		\mathbf{L}(T-h+1),\ldots,\mathbf{L}(T)
		\right)
		=
		\mathbf{z}(T)
		\equiv
		\begin{pmatrix}
			X(T)\\
			Y(T)
		\end{pmatrix}.
	\end{equation}
	The projection is constructed primarily from the recent evolution of $M_0$ and $M_1$, supplemented by statistics of the low-reliability tail; its explicit form is given in Methods. This construction maps the high-dimensional BP trajectory onto a two-dimensional diagnostic space while retaining information relevant to decoding convergence (Fig. \ref{fig:fig1}b).
	
	\begin{figure}
		\centering
		\includegraphics[width=\columnwidth]{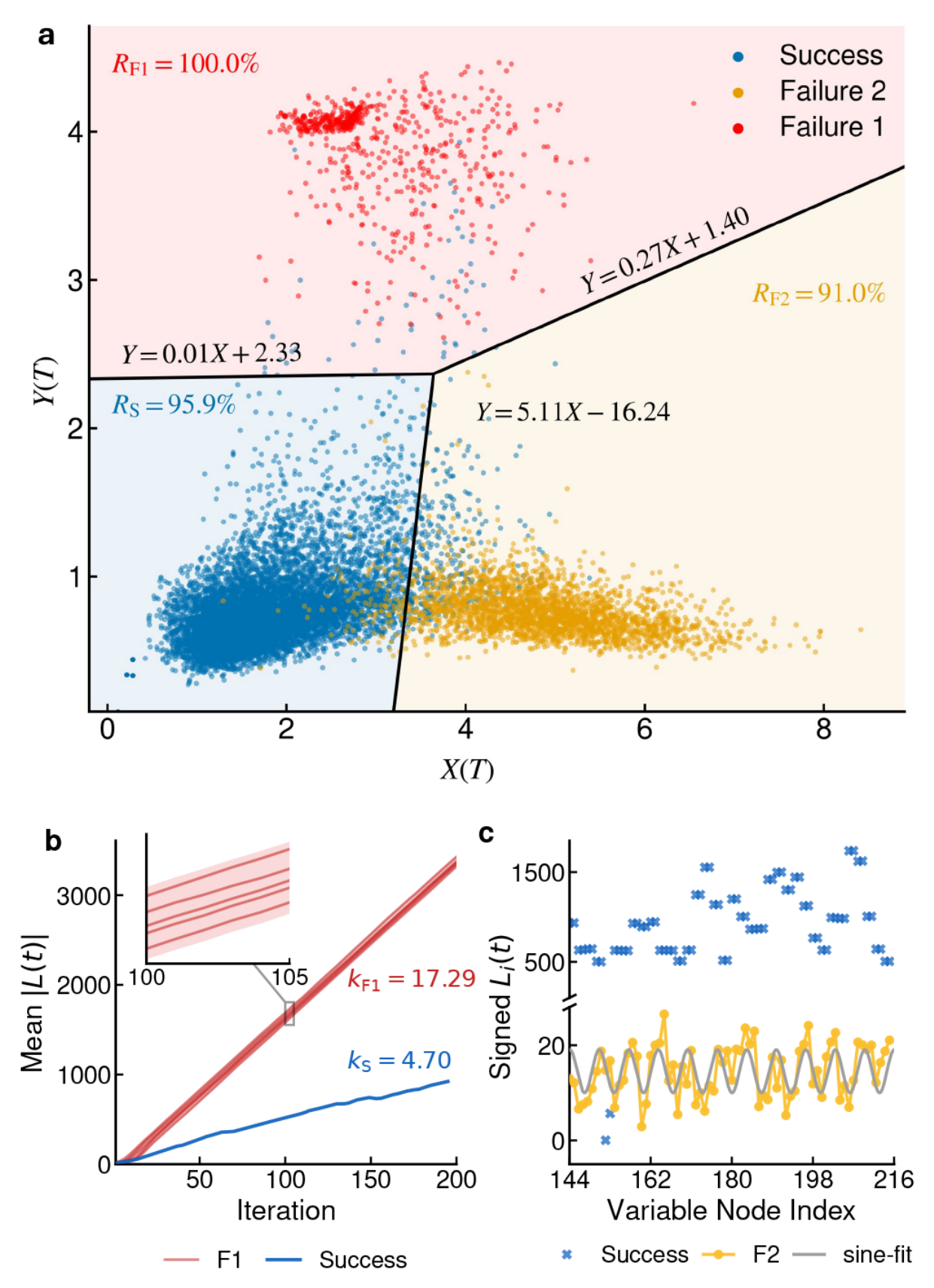}
		\caption{\textbf{Low-dimensional LLR representation separates distinct BP decoding regimes.}
			\textbf{a}, LDA partition of the two-dimensional representation
			$\mathbf{z}(T)=(X(T),Y(T))^{\mathsf T}$ constructed from low-confidence
			LLR statistics. Points denote successful, F1, and F2 decoding attempts,
			and black lines indicate the linear decision boundaries. The fractions of successful, F1 and F2 trajectories assigned
			to their respective decision regions are
			95.9\%, 100.0\%, and 91.0\%, respectively.
			\textbf{b}, Evolution of the mean absolute LLR for five
            representative F1 trajectories and a successful trajectory,
            showing continued confidence growth in F1 despite an unresolved
            syndrome. The inset magnifies the five F1 trajectories.
            The value $k_{\mathrm{F1}}=17.29$ denotes the mean of their
            individual slopes, whereas $k_{\mathrm{S}}=4.70$ is the slope
            of the successful trajectory.
			\textbf{c}, Final-iteration signed-LLR profiles for representative
			successful and F2 samples. The F2 sample exhibits a low-amplitude
			oscillatory pattern; the grey curve shows a sine fit.}
		\label{llr}
	\end{figure}
	
	The projected space reveals three characteristic classes of decoding trajectories (Fig.~\ref{llr}). Successful attempts occupy the region associated with a small population and severity of low-confidence variables. F1 trajectories retain a small set of unresolved variables while the overall LLR magnitude continues to increase, consistent with degeneracy-related locking. F2 trajectories are instead associated with low-amplitude oscillatory
	LLR patterns across variable nodes, consistent with trapping-set dynamics. We use linear discriminant analysis (LDA) to partition the projected space into three decision regions, $\mathbb{R}^{2}
	=
	\mathcal{R}_{\mathrm{S}}
	\cup
	\mathcal{R}_{\mathrm{F1}}
	\cup
	\mathcal{R}_{\mathrm{F2}}$,
	corresponding to successful, F1, and F2 trajectories, respectively,
	yielding a clear separation of the three dynamical regimes with
	consistently high correct-region ratios. The statistical motivation of the low-confidence observables is provided in
	Supplementary Information.
	
	The reduced representation also provides an online diagnostic of decoder evolution. By evaluating the projected state during decoding, we define the failure-detection stopping time as
	\begin{equation}
		T_{\mathrm{stop}}
		=
		\inf
		\left\{
		T:
		\mathbf{z}(T)
		\in
		\mathcal{R}_{\mathrm{F1}}
		\cup
		\mathcal{R}_{\mathrm{F2}}
		\right\}.
	\end{equation}
	Once a trajectory enters a region associated with persistent non-convergence, the corresponding decoding attempt is terminated before exhausting its full iteration budget. The dimensional reduction thus serves two complementary purposes: it provides an interpretable description of otherwise high-dimensional BP dynamics and converts this dynamical information into a practical early-termination criterion for real-time decoding.
	
	\subsection*{Dynamics-guided parallel decoding}
	
	We calibrate the message-weighting parameters of Min-Sum BP offline using simulated annealing\cite{sa} (Fig. \ref{fig:fig1}c); implementation details are given in
	Methods. Even with calibrated message passing, however, difficult
	decoding instances can require long serial BP trajectories, creating a latency tail that is incompatible with a fixed real-time decoding budget. We address this bottleneck by converting additional computational resources into parallel trajectory exploration rather than additional serial iterations.
	
	Specifically, we execute $N_{\mathrm{arm}}$ BP arms simultaneously (Fig. \ref{fig:fig1}d). Each arm, indexed by
$a=1,\ldots,N_{\mathrm{arm}}$, starts from an
independently perturbed LLR profile,
\begin{equation}
\mathbf{L}_{\mathrm{in}}^{(a)}
=
\mathbf{L}_{\mathrm{in}}+\boldsymbol{\epsilon}^{(a)},
\quad
\boldsymbol{\epsilon}^{(a)}
\sim\mathcal{N}(0,\sigma^2 \mathbf{I}),
\end{equation}
	where \(\mathbf{L}_{\mathrm{in}}\) is the posterior LLR vector obtained at the end of memory-augmented belief propagation (Mem-BP) preprocessing and \(\sigma\) controls the perturbation strength. The arms subsequently
	evolve under the same BP update rule but probe distinct nearby decoding
	trajectories. Because these trajectories are evaluated concurrently,
	increasing the parallel width does not increase the serial BP depth of
	an individual decoding attempt.
	
	The low-dimensional dynamical representation introduced above provides an
	online criterion for deciding which trajectories should continue. For arm
	$a$, the corresponding failure-detection stopping time is
	$T_{\rm stop}^{(a)}$, as defined above. An arm entering either failure
	region is terminated immediately, and its unused iteration budget is
	recycled to initialize a newly perturbed decoding trajectory. This
	terminate--reseed mechanism converts iterations that would otherwise be
	spent on unpromising trajectories into additional trajectory exploration.
	Because the arms are evaluated concurrently, this additional exploration
	does not extend the critical BP path.

	\begin{figure*}[t]
		\centering
		\includegraphics[width=\textwidth]{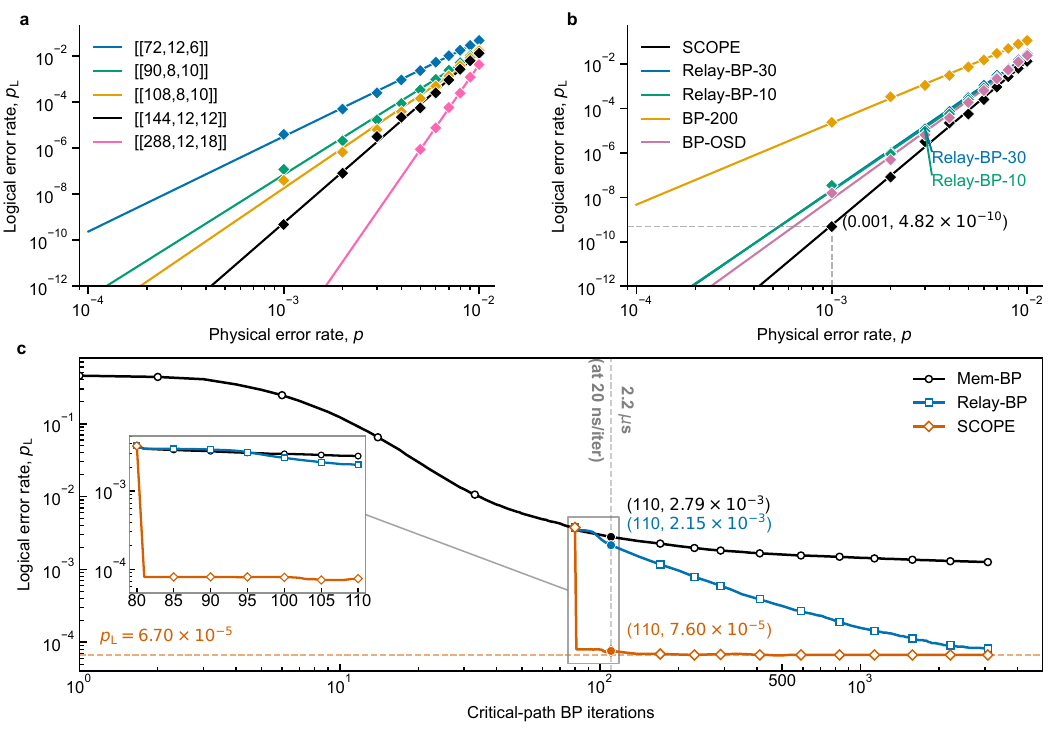}
		\caption{
			\textbf{Circuit-level decoding performance and convergence.}
			\textbf{a}, Logical error rate versus physical error rate for several
			BB codes decoded with SCOPE. For most data points, Monte Carlo sampling was continued until at least 100 logical failures were observed, yielding a relative standard error of approximately $10\%$ or less. Error bars are omitted for clarity.
			\textbf{b}, Comparison of SCOPE with BP-based decoders on the
			$[[144,12,12]]$ code. The benchmark configurations are SCOPE with 80 preprocessing iterations followed by four parallel arms of $120$ iterations each (200 critical-path BP iterations), pure BP with
			$200$ iterations, Relay-BP-10 with $80+10\times60$ iterations,
			Relay-BP-30 with $80+30\times20$ iterations, and BP-OSD with $10^4$
			BP iterations followed by OSD of combination-sweep (CS) order $7$. Here, the suffixes 10 and 30 denote the maximum numbers of post-preprocessing relay legs. In both \textbf{b} and \textbf{c}, Relay-BP stops upon finding the first syndrome-consistent solution or exhausting its BP iteration budget.
			The Relay-BP-10 and Relay-BP-30 curves nearly overlap over the plotted
			range.
			\textbf{c}, Convergence comparison on the $[[144,12,12]]$ code at $p=5\times10^{-3}$. The horizontal axis represents the critical-path BP iteration depth. SCOPE and Relay-BP share the same 80-iteration Mem-BP preprocessing stage; the inset enlarges the 80–110 iteration range. For SCOPE, each post-preprocessing point corresponds to a specified parallel-arm budget, with OSD applied only if all four arms remain unresolved after exhausting that budget. For Relay-BP, the curve beyond 80 iterations is generated by varying the maximum number of post-preprocessing relay legs, with a maximum of 30 BP iterations per leg. }
		\label{fig:performance-convergence}
	\end{figure*}

	\subsection*{Circuit-level decoding performance}
	
	We evaluate SCOPE under circuit-level noise using the decoding matrices
	and noise model provided by the \textsc{BivariateBicycleCodes}
	library~\cite{BravyiBivariateBicycleCodes}. Unless otherwise stated, the
	syndrome window contains $N_c=2$ measurement cycles, providing a short
	repeated-syndrome setting with time-like correlations.
	For a logical failure probability $P_L(N_c)$ over an $N_c$-cycle window,
	we report the equivalent logical error rate per syndrome cycle as
	\begin{equation}
		p_L
		=
		1-\left[1-P_L(N_c)\right]^{1/N_c}
		\simeq
		\frac{P_L(N_c)}{N_c},
	\end{equation}
	where the approximation applies in the low-error regime.
	
	For the numerical benchmarks below, SCOPE denotes the complete decoding
	pipeline. Each sample first undergoes an $80$-iteration Mem-BP
	preprocessing stage~\cite{muller2025}, followed by parity-check
	verification. Samples that remain unresolved are passed to four concurrent decoding
	arms. Within each arm budget, a trajectory identified as unpromising is
	terminated and reseeded using its remaining iterations. If the total
	post-preprocessing BP budgets of all four concurrent arms are exhausted
	without producing a valid correction, ordered-statistics decoding (OSD)~\cite{Roffe_2020} is applied as the
	final fallback stage. 
	
	Because the four decoding arms are executed concurrently, we characterize
	the real-time BP cost by the critical-path serial iteration depth. For a
	post-preprocessing budget of $B$ iterations per arm, this depth is
	$D_{\mathrm{serial}}=80+B$, rather than $80+4B$. This quantity describes
	the serial BP depth relevant to feedback latency and is distinct from the
	aggregate computation performed across the parallel arms. Unless
	otherwise stated, Figs.~\ref{fig:performance-convergence}a and
	\ref{fig:performance-convergence}b use $B=120$, corresponding to a
	critical-path depth of $80+120$ BP iterations.
	
	We first evaluate SCOPE across several BB codes with
	different block lengths and code distances
	(Fig.~\ref{fig:performance-convergence}a). Across all investigated code
	instances, the logical error rate is consistently suppressed as the
	physical error rate is reduced, showing that the observed decoding
	performance is not restricted to a single BB-code instance. The markers
	in Figs.~\ref{fig:performance-convergence}a and
	\ref{fig:performance-convergence}b denote direct Monte Carlo simulations,
	while the remaining portions of the curves are obtained from linear fits
	in the log--log domain.
	
	We next benchmark SCOPE against Relay-BP and BP-OSD on the
	$[[144,12,12]]$ BB code
	(Fig.~\ref{fig:performance-convergence}b). All decoders are evaluated
	under identical circuit-level noise, decoding matrices, and
	syndrome-window settings. Relay-BP and BP-OSD are simulated using the
	\textsc{Relay}~\cite{relaybpSoftware2026} and
	\textsc{Ldpc}~\cite{roffe2022ldpc} implementations, respectively. Under
	these matched conditions, SCOPE achieves a lower logical error rate
	across the full investigated physical-error range while operating at a
	substantially smaller critical-path BP depth. In particular, direct
	Monte Carlo simulation at $p=10^{-3}$ gives
	$p_L=4.82\times10^{-10}$, approximately $70$ times lower than Relay-BP
	and $30$ times lower than BP-OSD under the corresponding benchmark
	configurations.
	
	We further compare decoding performance as a function of the critical-path BP iteration budget $D_{\rm serial}$ on the $[[144,12,12]]$ code at
	$p=5\times10^{-3}$ (Fig.~\ref{fig:performance-convergence}c). SCOPE converges more rapidly than Mem-BP
	and Relay-BP, reaching substantially lower logical error rates within the
	same serial iteration budget and approaching
	$p_L=6.70\times10^{-5}$ at larger post-preprocessing budgets.
	
	Finally, we relate the critical-path BP depth to a representative hardware timing budget. Taking a QEC cycle time of
	$1.1~\mu\mathrm{s}$ reported in Ref.~\cite{Google2024}, the $N_c=2$
	syndrome window spans approximately
	$2.2~\mu\mathrm{s}$. Using a representative BP iteration time of
	$20~\mathrm{ns}$ assumed in the hardware-timing estimate of Ref.~\cite{muller2025}, this interval corresponds to approximately
	$110$ serial BP iterations. At the corresponding $80+30=110$
	critical-path operating point, SCOPE achieves
	$p_L=7.60\times10^{-5}$, lower than those of Relay-BP and Mem-BP by factors of approximately 28 and 37, respectively, under the same BP iteration budget. Thus, SCOPE retains its accuracy advantage under a hardware-motivated constraint on serial BP depth.
	
	To assess how the covariant structure translates into hardware performance, we evaluate a register-transfer-level (RTL) prototype of the Mem-BP core that runs two perturbed decoding trajectories in parallel. Each arm uses four cooperating
	compute partitions and maintains independent message and belief
	states. The first syndrome-consistent result is selected and the other trajectory is cancelled. For one CSS sector of the \([[144,12,12]]\) code with \(N_c=2\), we evaluate 4096 samples from the circuit-level noise model at \(p=5\times10^{-3}\). At a \(250\,\mathrm{MHz}\) simulation clock, the prototype yields
	a mean input-to-result latency of \(173.8\,\mathrm{ns}\) and a
	99th-percentile latency of \(732\,\mathrm{ns}\). On the
	same sample bank, the perturbed two-arm configuration reduces
	the mean latency by approximately \(13\%\) relative to the
	unperturbed single-trajectory implementation
	(\(199.6\,\mathrm{ns}\)). All selected correction vectors match an independent fixed-point reference. These results provide RTL-level evidence for the practical benefit of the multi-trajectory strategy within the covariance-structured framework, translating concurrent trajectory exploration into a measurable latency advantage. The prototype achieves a nanosecond-scale mean response, with \(99\%\) of measured responses below one microsecond. Implementation details, verification procedures and complete latency statistics are provided in Supplementary Information.

	\subsection*{Implications for real-time quantum decoding}
	
	Our results suggest that real-time quantum decoding can be viewed as a
	structured dynamical inference process, in which both the structure of
	the decoding problem and the evolution of the decoder can be exploited
	to control computation. The different ingredients of this framework
	have complementary ranges of applicability. Temporal covariance follows
	from repeated syndrome extraction and is therefore not intrinsically
	restricted to a particular code family, whereas additional spatial
	covariance depends on the symmetries of the underlying code and circuit.
	By contrast, the low-dimensional dynamical representation relies on the
	evolution of BP beliefs rather than on BB-code-specific structure, and
	therefore provides a basis for extension to other qLDPC families decoded by
	BP or related message-passing algorithms. The dynamics-guided parallel
	architecture inherits this decoder-level generality.
	
	The present numerical and RTL demonstrations focus on BB codes
	with \(N_c=2\) syndrome cycles. The two-arm RTL prototype
	demonstrates concurrent trajectory execution within the
	covariance-structured Mem-BP core, providing an implementation
	basis for further hardware development. Extending the framework
	to longer syndrome histories and broader qLDPC families remains
	an important direction~\cite{Skoric_2023,turner2026}.
	SCOPE trades parallel computational width for reduced serial
	BP depth, making the hardware resource--latency trade-off
	central to further FPGA or ASIC implementations. The
	covariance-based partitioning also provides a basis for future
	distributed implementations across multiple RFSoC devices.
	Building on the fixed-point core, further work will evaluate
	the complete decoding pipeline on physical hardware and
	quantify how trajectory control, post-processing and
	communication overhead affect end-to-end feedback
	latency~\cite{maurya2026,ye2026}. Together, these results connect
	circuit structure, decoder dynamics and parallel execution,
	providing a route towards accurate qLDPC decoding under
	real-time constraints.
	
	\newpage
	\section*{Methods}
	
	\subsection*{Circuit-level fault model and decoding-matrix construction}
	
	We construct the circuit-level decoding model for the \(Z\)-error sector of the BB code~\cite{BravyiBivariateBicycleCodes}. For each BB code, the stabilizer
	matrices specify the connectivity between data qubits and check ancillas.
	Using this connectivity, we construct the fixed syndrome-extraction circuit
	and repeat it for the prescribed number of measurement cycles. Elementary
	circuit faults are then introduced individually at the relevant preparation,
	measurement, and two-qubit-gate locations and propagated through the
	repeated circuit.
	
	For each elementary fault, we record the complete history of the relevant
	check-measurement outcomes. Detector events are obtained from differences between consecutive
	measurements of the same check, such that a detector at round $t$ records
	$m_t\oplus m_{t-1}$, where $m_t\in\mathbb{F}_2$ denotes the measured
	syndrome outcome of that check in round $t$. After the $N_c$ noisy
	syndrome-extraction cycles, two additional noiseless cycles are appended to
	close the finite decoding window and capture the terminal syndrome response.
	The propagated data-qubit state is also tested against the corresponding
	logical operators, yielding the logical signature associated with each
	circuit fault. Each microscopic fault is therefore characterized by
	both its detector history and its logical action.
	
	Microscopic faults producing identical detector and logical signatures are
	grouped into a single effective fault, with their associated probabilities
	combined into one prior. The detector components of these effective
	signatures form the decoding matrix $D$, while their logical components form
	a separate logical-response matrix $A$. The resulting circuit-level model is
	therefore specified by $(D,A,\mathbf{p})$, where
	$\mathbf{p}=(p_1,\ldots,p_N)$ contains the prior probabilities of the
	effective circuit faults. Monte Carlo samples are generated by independently
	drawing these effective fault variables according to their priors and
	evaluating the corresponding detector syndrome through $D$. This sampling defines an independent effective-fault approximation to the underlying circuit noise model. Thus, the
	variables decoded by BP represent effective circuit-fault classes rather
	than individual physical data-qubit errors.
	
	\subsection*{Belief-propagation decoding and parameter calibration}
	
	Given the circuit-level model $(D,A,\mathbf{p})$ constructed above, the
	decoding pipeline employs two successive belief-propagation stages. The
	initial preprocessing stage uses Mem-BP~\cite{muller2025}, which
	introduces memory by feeding the posterior belief from the preceding
	iteration back into the variable-node bias. In the notation used below,
	the fixed prior contribution $L(p_i)$ is replaced during Mem-BP by
	$(1-\gamma_i)L(p_i)+\gamma_i L_i(t-1)$, where the memory strength is uniform across variable nodes,
\(\gamma_i=\gamma\) for all \(i\). The subsequent parallel decoding stage uses the
	Normalized Min-Sum Algorithm (NMSA)\cite{nmsa, minsum} without this memory feedback.
	
	For variable node $v_i$ with prior fault probability $p_i$, the intrinsic
	log-likelihood ratio (LLR) is
	$L(p_i)=\log[(1-p_i)/p_i]$. We denote by $q_{ij}^{(t)}$ the message from
	variable node $v_i$ to check node $c_j$ at iteration $t$, and by
	$r_{ji}^{(t)}$ the corresponding check-to-variable message. The messages
	are initialized as $q_{ij}^{(0)}=L(p_i)$. The check-node update is
	
	\begin{equation}
		\begin{aligned}
			r_{ji}^{(t)}
			&=
			\alpha (-1)^{s_j}
			\prod_{i'\in\mathcal N(c_j)\setminus v_i}
			\operatorname{sgn}\!\left(q_{i'j}^{(t-1)}\right)
			\\
			&\quad \times
			\min_{i'\in\mathcal N(c_j)\setminus v_i}
			\left|q_{i'j}^{(t-1)}\right|,
		\end{aligned}
	\end{equation}
	where $\alpha$ rescales the Min-Sum check-node message. The
	variable-to-check update is
	
	\begin{equation}
		q_{ij}^{(t)}
		=
		L(p_i)
		+
		\eta
		\sum_{j'\in\mathcal N(v_i)\setminus c_j}
		r_{j'i}^{(t)},
	\end{equation}
	
	where $\eta$ controls the contribution of the accumulated extrinsic
	check-node messages. The posterior LLR associated with variable node $v_i$
	after iteration $t$ is
	
	\begin{equation}
		L_i(t)
		=
		L(p_i)
		+
		\eta
		\sum_{j\in\mathcal N(v_i)}
		r_{ji}^{(t)} ,    
	\end{equation}
	and the posterior-belief vector is denoted by
	$\mathbf{L}(t)=(L_1(t),\ldots,L_N(t))$. A hard estimate is obtained as
	$\hat{\xi}_i^{(t)}=\mathbf{1}[L_i(t)<0]$.
	
	For a sampled circuit-fault configuration $\boldsymbol{\xi}$, the measured
	detector syndrome is $\mathbf{s}=D\boldsymbol{\xi}$. During decoding, a
	candidate estimate is syndrome-consistent when
	$D\hat{\boldsymbol{\xi}}^{(t)}=\mathbf{s}$. In the numerical simulations,
	a syndrome-consistent estimate is counted as logically successful only if
	the residual fault configuration has trivial logical action, namely
	
	\begin{equation}
		D\hat{\boldsymbol{\xi}}^{(t)}=\mathbf{s},
		\qquad
		A\!\left(
		\boldsymbol{\xi}+\hat{\boldsymbol{\xi}}^{(t)}
		\right)=\mathbf{0},   
	\end{equation}
	
	where addition is over $\mathbb F_2$. Equivalently, the second condition
	requires
	$A\hat{\boldsymbol{\xi}}^{(t)}=A\boldsymbol{\xi}$.
	
	The decoding performance of NMSA depends sensitively on the two
	message-weighting parameters $\alpha$ and $\eta$. For the real-valued update rules considered here, their effects on the posterior beliefs are coupled through the product \(\alpha\cdot\eta\). We nevertheless retain \(\alpha\) and \(\eta\) as separate implementation parameters and calibrate them jointly using simulated annealing (SA). For each candidate pair $(\alpha,\eta)$, we evaluate the
	logical failure probability under a fixed decoding configuration and
	formulate the calibration problem as
	
	\[
	(\alpha^\star,\eta^\star)
	\in
	\underset{\alpha,\eta}{\arg\min}\;
	P_L(\alpha,\eta).
	\]
	
	Because $P_L(\alpha,\eta)$ is estimated from finite Monte Carlo samples, the
	resulting objective is statistically noisy and need not vary smoothly over
	the parameter space. SA is therefore used to explore the parameter space efficiently while avoiding exhaustive sweeps. At each
	annealing step, a nearby candidate $(\alpha',\eta')$ is proposed and
	accepted according to the standard Metropolis rule. The optimized
	parameters are fixed after calibration and used throughout the subsequent
	decoding simulations.
	
	\subsection*{Covariance-based layered implementation}

	For the BB circuit models considered here, spatiotemporal
	covariance allows repeated detector-support patterns in
	\(D\) to be represented by shared templates.
	These repeated connections support parallel node updates
	across translated regions. We partition the graph in both
	detector time and space to organize this parallel computation
	and limit the BP messages exchanged between partitions.
	The partitions jointly execute a single BP trajectory.
	We use \(u\) to label detector layers and retain \(t\) for
	the BP iteration index. Check node \(c_j\) has coordinates
	\((u_j,\beta_j)\), where
	\(\beta_j\in\mathcal M=\mathbb Z_\ell\times\mathbb Z_m\).
	The finite detector history contains the \(N_c\) noisy
	syndrome-extraction cycles and the two noiseless closing
	cycles defined above, with layers indexed by
	\(u=0,\ldots,N_c+1\).
	
	We first group the nonzero column supports of \(D\) into
	spatial translation orbits. Within each orbit, a reference
	support is selected by lexicographically ordering the
	detector coordinates, first by layer and then by spatial
	position. Subtracting the earliest detector layer of the
	reference support gives a relative spatiotemporal pattern.
	Identical patterns, including those arising at different
	temporal anchors, are stored as a shared template
	\(\mathcal S_\kappa
	\subset\mathbb Z_{\geq0}\times\mathcal M\).
	Each variable \(v_i\) with nonzero detector support is then
	specified by a template index \(\kappa_i\), a temporal
	anchor \(u_i\), and a spatial translation \(g_i\), such that
	\begin{equation}
	D_{ji}
	=
	\mathbf{1}\!\left[
	\bigl(u_j-u_i,\beta_j-g_i\bigr)
	\in\mathcal S_{\kappa_i}
	\right].
	\label{eq:layered_template_reconstruction}
	\end{equation}
	Here, \(u_i\) is the earliest detector-response layer of
	column \(i\), and \(g_i\) translates the reference support
	to that column. The temporal anchor is a coordinate of
	the detector response and need not coincide with the
	microscopic fault time. Spatial differences are evaluated
	in \(\mathcal M\), while detector time remains a finite,
	nonperiodic coordinate.
	
	The templates and their allowed anchors are extracted from
	the complete finite-window matrix.
	Equation~\eqref{eq:layered_template_reconstruction}
	therefore retains the actual initial and terminal supports,
	including the closing layers. Fault priors remain attached
	to their original variables, so copies of the same template
	at different temporal anchors can have different priors.
	The logical-response matrix \(A\) and the identity of each
	effective-fault variable are also preserved. Template
	sharing provides a compact description of connectivity;
	each variable retains its own posterior belief, and each
	expanded Tanner edge carries independent messages.
	
	In the present implementation, consecutive detector layers
	are grouped into temporal intervals, and each interval is
	subdivided into spatial regions. Each resulting region is
	assigned to a compute partition.
	Let \(\Pi(u,g)\in\{1,\ldots,P\}\) denote the partition
	assigned to coordinate \((u,g)\). The owners of variable
	and check nodes are
	\begin{equation}
	\begin{aligned}
	\pi_{\mathrm V}(i) &= \Pi(u_i,g_i),\\
	\pi_{\mathrm C}(j) &= \Pi(u_j,\beta_j).
	\end{aligned}
	\label{eq:layered_node_ownership}
	\end{equation}
	The corresponding index sets are
	\(\mathcal V_b=\{i:\pi_{\mathrm V}(i)=b\}\) and
	\(\mathcal C_b=\{j:\pi_{\mathrm C}(j)=b\}\).
	Thus, every variable and every check has a unique owner.
	Each partition maintains the posterior beliefs and
	outgoing variable-to-check messages of its variables,
	the outgoing check-to-variable messages of its checks,
	and buffers for the incoming messages required by these
	nodes.
	
	For the \([[144,12,12]]\) implementation with
	\((\ell,m)=(12,6)\) and \(N_c=2\), detector layers are
	grouped into \(\{0\}\) and \(\{1,2,3\}\).
	Within each temporal group, the spatial lattice is divided
	into \(0\leq x<6\) and \(6\leq x<12\), with all
	\(y\) coordinates retained in each region.
	The two temporal groups and two spatial regions form a
	\(2\times2\) layout of four cooperating on-chip partitions
	per decoding trajectory. The temporal split supplies part
	of the parallelism, allowing four compute partitions to use
	only two spatial regions. Each partition contains 36 spatial
	anchors, allowing repeated local node circuits to be
	instantiated in parallel.
	All connections across spatial and temporal partition
	boundaries are retained, so the partitions jointly update
	one fault estimate.
	
	Let \(D^{bb'}\) denote the restriction of \(D\) to rows
	\(\mathcal C_b\) and columns \(\mathcal V_{b'}\).
	The associated edge sets and the set of crossing edges are
	\begin{equation}
	\begin{aligned}
	\mathcal E_{bb'}
	&=
	\left\{
	(i,j):
	i\in\mathcal V_{b'},\
	j\in\mathcal C_b,\
	D_{ji}=1
	\right\},\\
	\mathcal E_{\partial}
	&=
	\bigcup_{b\ne b'}\mathcal E_{bb'}.
	\end{aligned}
	\label{eq:layered_boundary_edges}
	\end{equation}
	For an edge in \(\mathcal E_{bb'}\), the message
	\(q_{ij}\) is computed by partition \(b'\) and delivered
	to partition \(b\), whereas \(r_{ji}\) is computed by
	partition \(b\) and delivered to partition \(b'\).
	When \(b=b'\), both transfers are local. The routing
	tables are generated from the template-expanded graph.
	All blocks \(D^{bb'}\) participate in the same decoding
	problem, and both message directions are retained on
	edges connecting different detector layers.
	
	The communication benefit can be quantified directly from
	\(D\). For this window, the implemented layout has 1968
	crossing edges out of 5472 Tanner edges, with the remaining
	3504 edges local to their partitions. Purely spatial four-way
	divisions yield 2640 crossing edges for four equal strips
	along \(x\), and 2544 for two equal regions in each spatial
	direction. Both spatial layouts retain all detector layers
	in each region and use the same variable-anchor ownership
	rule. Each crossing edge carries one \(q_{ij}\) and one
	\(r_{ji}\) message per complete BP iteration. The implemented
	layout therefore exchanges approximately \(25\%\) and
	\(23\%\) fewer BP message values between partitions than
	these two layouts, respectively.
	
	For software verification, the partitioned decoder is
	evaluated using a synchronous floating-point implementation
	with a flooding schedule and the complete syndrome window
	available before iteration.
	After initialization for the relevant BP stage, variable
	owners deliver \(q_{ij}^{(0)}\) and the initial hard
	estimates to the check owners. At iteration \(t\), all
	check nodes use the same generation of messages
	\(q_{ij}^{(t-1)}\) to compute \(r_{ji}^{(t)}\) according
	to the check-node rule defined above. The resulting
	messages are delivered to the variable owners, and a
	synchronization barrier ensures that all required
	check-to-variable messages have arrived before the
	variable-node updates begin. Each variable owner then
	updates \(L_i(t)\), \(q_{ij}^{(t)}\), and the hard
	estimate \(\hat{\xi}_i^{(t)}\). For Mem-BP, the memory
	term uses the preceding posterior \(L_i(t-1)\), which
	is retained by the same variable owner. The updated
	variable-to-check messages and hard estimates are
	delivered to the check owners, and a second synchronization
	barrier precedes syndrome verification and the next
	iteration.
	
	Under this reference schedule, the partitioned execution
	preserves the dependencies of the unpartitioned BP update.
	For matched initialization and update parameters, each
	check receives the same incoming messages as in the complete
	graph, and each variable subsequently receives the same
	complete set of updated check messages. Applying this
	argument at successive iterations establishes equivalence
	of the partitioned and unpartitioned trajectories in exact
	arithmetic. Numerical verification additionally matches
	the arithmetic, saturation rules and accumulation order.
	
	Global syndrome consistency is evaluated from local
	check residuals. Writing \(s_{\mathcal C_b}\) and
	\(\hat{\xi}_{\mathcal V_b}^{(t)}\) for the restrictions
	of the syndrome and hard estimate to the corresponding
	index sets, each check owner computes
	\begin{equation}
	\begin{aligned}
	d_b(t)
	&=
	s_{\mathcal C_b}
	\oplus
	\bigoplus_{b'=1}^{P}
	D^{bb'}\hat{\xi}_{\mathcal V_{b'}}^{(t)},\\
	n_{\mathrm{unsat}}(t)
	&=
	\sum_{b=1}^{P}\operatorname{wt}\!\left[d_b(t)\right].
	\end{aligned}
	\label{eq:layered_syndrome_reduction}
	\end{equation}
	The matrix products and \(\oplus\) are over
	\(\mathbb F_2\), and \(\operatorname{wt}\) denotes
	Hamming weight. All local residuals are evaluated from
	the same candidate \(\hat{\xi}^{(t)}\) using the detector
	values of the complete decoded window.
	Summing their Hamming weights determines global syndrome
	consistency: \(n_{\mathrm{unsat}}(t)=0\) if and only if
	\(D\hat{\xi}^{(t)}=s\).
	In the synchronous software reference, this condition is
	checked after initialization and after each complete
	iteration.
	
	The predicted logical action can likewise be assembled
	from the variable owners. Let \(A_b\) contain the columns
	of \(A\) indexed by \(\mathcal V_b\). Then
	\begin{equation}
	A\hat{\xi}^{(t)}
	=
	\bigoplus_{b=1}^{P}
	A_b\hat{\xi}_{\mathcal V_b}^{(t)}.
	\label{eq:layered_logical_reduction}
	\end{equation}
	Each partition computes its local contribution, and the
	contributions are combined by XOR. The sampled fault
	configuration \(\xi\) is used only in the evaluation of
	logical success defined above; the convergence decision
	uses the measured syndrome and the current estimate.
	
	We verify the partitioning using a double-precision Python
	implementation with both an in-process backend and a
	multiprocessing backend with persistent partition states.
	An independent reference decoder operates directly on
	the complete matrix \(D\). The Python comparisons use
	matched update parameters and symmetric LLR saturation
	at magnitude \(100\), with the original edge accumulation
	order retained. For a four-partition software test with
	\(N_c=2\) and \(p=5\times10^{-3}\), the first
	32 samples of a 4096-sample run were compared with the
	reference at every iteration. All variable-to-check
	messages, check-to-variable messages and posterior beliefs
	agreed with zero measured absolute difference, and the
	hard estimates and stopping iterations also matched.
	The same 32 samples were verified with the multiprocessing
	backend. Additional tests cover alternative spatial and
	temporal partitions, finite-window boundaries and the
	completeness of boundary exchanges.
	
	The register-transfer-level (RTL) prototype implements
	two perturbed Mem-BP trajectories for the same syndrome
	window, with four on-chip compute partitions per arm.
	Each arm starts directly from its own quantized,
	perturbed physical-prior LLRs, which remain fixed during
	decoding, and is allocated a budget of 80 iterations.
	Messages use a signed five-bit representation with one
	fractional bit, with coefficients \(\alpha=\eta=1\)
	and \(\gamma=26/256\).
	Two check-node stages followed by two variable-node
	stages give a four-cycle iteration interval per arm,
	corresponding to \(16\,\mathrm{ns}\) at the
	\(250\,\mathrm{MHz}\) simulation clock.
	Each iteration uses messages stored by the preceding
	iteration, and syndrome verification overlaps the
	following check-node calculation.
	
	The first syndrome-consistent arm supplies the selected
	result; arm zero wins simultaneous ties, and the other
	arm is cancelled. If both arms exhaust their budgets,
	the output is marked as failed and arm zero's final
	estimate is retained for diagnosis.
	Across the 4096 RTL test windows, the returned fault
	estimates, iteration counts, convergence flags and logical
	outputs matched an independent fixed-point reference.
	
	The reported hardware timings are obtained from RTL
	functional simulation and cover this two-arm Mem-BP
	core together with input assembly and result registration.
	Input-to-result latency is measured from acceptance of
	the first syndrome word to assertion of result valid,
	under continuous input and no output backpressure,
	and includes windows that exhaust the iteration budget.
	The preprocessing-to-NMSA sequence, online dynamical
	classification, failure-triggered reseeding and OSD of
	the complete software pipeline are outside this timed
	path. Fixed-point update rules and timing details are
	provided in Supplementary Information.
	
	The same four-partition organization provides a route to
	distributed decoding by assigning its partitions to separate
	RFSoCs or FPGAs. Each device stores and updates its assigned
	variables and checks, while crossing-edge messages connect
	the devices and local contributions are combined for global
	syndrome and logical-output checks. The smaller crossing-edge
	set reduces the amount of BP data exchanged between devices
	per iteration.
	Preserving a decoding trajectory across devices requires
	the same priors and arithmetic rules for that trajectory,
	together with consistent window, arm and iteration
	identities for exchanged messages.
	The execution schedule must account for communication
	delays while preserving these dependencies.
	The graph partitioning can be reused across arms, while
	each arm retains independent perturbed priors, messages
	and belief states.

	\subsection*{Low-dimensional representation of BP dynamics}
	
	To characterize the evolution of BP beyond its final hard decision, we
	construct a low-dimensional representation from the posterior LLR state
	$\mathbf{L}(t)$ defined above. Writing
	$\ell_i(t)=|L_i(t)|$ for the reliability of variable node $i$, the
	low-confidence population is quantified by
	$M_0(t)=\sum_i \mathbf{1}[\ell_i(t)<\tau]$, where $\tau$ is a fixed
	reliability threshold. In the numerical implementation, we use the
	five-iteration moving average
	$\widetilde M_0(T)=\frac{1}{5}\sum_{u=0}^{4}M_0(T-u)$ to suppress
	iteration-to-iteration fluctuations.
	
	The severity of the low-confidence tail is represented by a practical
	surrogate of the first low-confidence moment introduced in the main text.
	Let
	$\ell_{(1)}(T)\leq \ell_{(2)}(T)\leq\cdots\leq\ell_{(N)}(T)$
	denote the ordered reliabilities. We define
	$\widetilde M_1(T)=\frac{1}{5}\sum_{j=1}^{5}\ell_{(j)}(T)$ as the mean
	reliability of the five least reliable variables. Although
	$\widetilde M_1$ has the opposite monotonic orientation to the theoretical
	first moment $M_1$---more severe low-confidence defects correspond to
	larger $M_1$ but smaller $\widetilde M_1$---both quantify the severity of
	the same low-reliability tail. We further use $k_{10}(T)$, the fitted
	growth slope of the low-reliability tail over the preceding ten BP
	iterations, and $q_{5\%}(T)$, the fifth percentile of the reliability
	distribution.
	
	The two diagnostic coordinates used for trajectory classification are
	
	\begin{equation}
		\begin{aligned}
			X(T)
			&= \log_{10}\!\left[1+\widetilde M_0(T)\right] \\
			&\quad + \alpha_X \log_{10}\!\left[
			1+\frac{1}{\widetilde M_1(T)+\epsilon}
			\right] \\
			&\quad + \beta_X \log_{10}\!\left[
			1+\frac{1}{\max(k_{10}(T),0)+\delta}
			\right], \\[2mm]
			Y(T)
			&= \log_{10}\!\left[
			\frac{1}{\widetilde M_1(T)+1}
			\right] \\
			&\quad + \gamma_Y
			\log_{10}\!\left[q_{5\%}(T)+1\right].
		\end{aligned}
	\end{equation}
	
	The projected decoder state is denoted by
	$\mathbf{z}(T)=(X(T),Y(T))^{\mathsf T}$. Here,
	$\alpha_X$, $\beta_X$, and $\gamma_Y$ set the relative weights of the
	different low-tail statistics, while $\epsilon$ and $\delta$ are small
	positive regularization constants introduced to avoid singular
	denominators.
	
	Linear discriminant analysis (LDA) is then fitted to labeled decoding
	trajectories in the $(X,Y)$ plane, defining three decision regions
	$\mathcal R_S$, $\mathcal R_{F1}$, and $\mathcal R_{F2}$ corresponding to
	successful, F1, and F2 trajectories, respectively. Once the projection
	parameters and LDA boundaries are determined, they are kept fixed during
	the decoding benchmarks.
	
	During BP evolution, $\mathbf{z}(T)$ is evaluated repeatedly as the
	posterior LLR trajectory evolves. The first entrance into either failure
	region defines the failure-detection stopping time
	
	\[
	T_{\mathrm{stop}}
	=
	\inf\left\{
	T:
	\mathbf{z}(T)\in
	\mathcal R_{F1}\cup\mathcal R_{F2}
	\right\}.
	\]
	
	A trajectory satisfying this condition is marked for early termination,
	whereas trajectories that have not entered either failure region remain
	active. 
	
	\subsection*{Dynamics-guided parallel decoding architecture}
	
	Samples that remain unresolved after the initial BP preprocessing and
	syndrome verification are passed to a parallel ensemble of decoding
	trajectories. Let $\mathbf{L}_{\rm in}$ denote the common LLR profile
	entering this stage. Each arm is initialized from an independent
	perturbation of $\mathbf{L}_{\rm in}$, as defined in the main text, and
	subsequently evolves under the same NMSA update rule. The arms are executed
	concurrently, so increasing the ensemble width enlarges the set of explored
	trajectories without increasing the serial iteration depth of an individual
	arm.
	
	During decoding, each active arm is continuously monitored using the
	low-dimensional representation introduced above. At iteration $T$, the
	posterior LLR history of arm $a$ is mapped to
	$\mathbf{z}^{(a)}(T)$ and tested against the fixed decision regions. If the
	trajectory first enters either $\mathcal{R}_{F1}$ or
	$\mathcal{R}_{F2}$, it is terminated at $T_{\rm stop}^{(a)}$. An arm that
	has not entered either failure region remains active until it produces a
	syndrome-consistent estimate, $D\hat{\boldsymbol{\xi}}=\mathbf{s}$, or
	exhausts its available iteration budget.
	
	Early termination releases the unused iteration budget of an unpromising
	trajectory. If arm $a$ is launched with an iteration budget $B^{(a)}$ and
	is terminated after $n_{\rm stop}^{(a)}$ iterations, the remaining budget is
	
	\[
	B_{\rm rem}^{(a)}
	=
	B^{(a)}-n_{\rm stop}^{(a)} .
	\]
	
	This remaining budget is used to restart the same arm with a newly perturbed trajectory. A fresh independent perturbation drawn from
	the same distribution is applied to the common input LLR profile, and the
	newly initialized trajectory is subjected to the same BP evolution and online
	dynamical monitoring. Thus, iterations released by trajectories already
	identified as unpromising are converted into additional independent
	trajectory exploration. The terminate–reseed procedure is repeated while the arm's iteration budget remains and no syndrome-consistent correction has been found by any arm.
	
	In the complete decoding pipeline, a fixed Mem-BP preprocessing stage is
	followed by syndrome verification before the dynamics-guided parallel
	stage. If the available BP budget is exhausted without producing a valid
	correction, ordered-statistics decoding (OSD) is invoked as the final
	fallback.
	
\section*{Acknowledgments}
We thank Hao Song, Zongpei Yue, Xinwei Li for their insightful discussions. This work was supported by the National Natural Science
Foundation of China (NSFC; grant nos. 92565306, 92476205, 92476001 and 92265205), the National Key Research and Development Program
of China (grant no. 2025YFA1411600),
and the Quantum Science and Technology--National
Science and Technology Major Project
(grant no. 2021ZD0302104).

	\bibliography{references_v1}

\end{document}